# Terahertz imaging and spectroscopy of large-area single-layer graphene

J. L. Tomaino,[1] A. D. Jameson,[1] J. W. Kevek,[1] M. J. Paul,[1] A. M. van der Zande,[2] R. A. Barton,[3] P. L. McEuen,[2,4] E. D. Minot,[1] and Yun-Shik Lee[1,*]

[1]Department of Physics, Oregon State University, Corvallis, OR 97331, USA
[2]Laboratory of Atomic and Solid-State Physics, Cornell University, Ithaca, NY 14853, USA
[3]School of Applied and Engineering Physics, Cornell University, Ithaca, NY 14853, USA
[4]Kavli Institute at Cornell for Nanoscale Science, Cornell University, Ithaca, NY 14853, USA
[*]leeys@physics.oregonstate.edu

**Abstract:** We demonstrate terahertz (THz) imaging and spectroscopy of a $15\times15$-mm$^2$ single-layer graphene film on Si using broadband THz pulses. The THz images clearly map out the THz carrier dynamics of the graphene-on-Si sample, allowing us to measure sheet conductivity with sub-mm resolution without fabricating electrodes. The THz carrier dynamics are dominated by intraband transitions and the THz-induced electron motion is characterized by a flat spectral response. A theoretical analysis based on the Fresnel coefficients for a metallic thin film shows that the local sheet conductivity varies across the sample from $\sigma_s = 1.7\times10^{-3}$ to $2.4\times10^{-3}$ $\Omega^{-1}$ (sheet resistance, $\rho_s = 420 - 590$ $\Omega$/sq).

## 1. Introduction

Graphene is composed of a single-atom-thick layer of carbon atoms arranged in a two-dimensional honeycomb lattice [1]. The unique electronic structure of graphene gives rise to massless charge carriers and ballistic transport on a submicron scale at room temperature [2,3]. The exceptional electronic properties of graphene have sparked intensive research into futuristic applications ranging from nanometer-scale switches to single molecule detection [4-9]. In particular, the high electron mobility of graphene points to great potential for broadband communications and high-speed electronics operating at terahertz (THz) switching rates [10-12]. Practical device applications require large-area graphene films, therefore, there is great interest in optimizing the growth of high-quality graphene films [13-15] and probing the electronic properties of these films at ultrafast time scales. This interest motivates our current measurements of large-area graphene by THz imaging and time-domain spectroscopy.

So far, two methods of graphene fabrication have shown promising results for scalable production; (i) epitaxial growth on SiC substrates [13] and (ii) chemical vapor deposition (CVD) on metal (Ni or Cu) layers [14,15]. Epitaxial graphene on SiC has been studied by THz spectroscopy which yielded an estimate of carrier scattering time ~ 2 fs [16]. Growth of graphene by CVD onto Cu foil yields large-area graphene that is > 90% single layer [15] and has recently been scaled up to the meter scale [17]. After wet transfer of graphene from Cu to a device substrate, typical carrier sheet density is $n_{2d}$ ~ $4 \times 10^{12}$ cm$^{-2}$ [17]. Reported sheet resistances of such single layer graphene vary widely, for example $\rho_s$ = 150 Ω/sq [17] versus 510 Ω/sq [18], corresponding to effective mobilities, $\mu = (\rho_s n_{2d} e)^{-1}$ ~ 10,000 cm$^2$s$^{-1}$V$^{-1}$ and 3,000 cm$^2$s$^{-1}$V$^{-1}$ respectively.

In this paper, we present the first THz imaging and time-domain spectroscopy (TDS) of large area, single-layer graphene that is grown on Cu-foil and subsequently transferred to a substrate. We have measured the transmission of a pulsed, spatially-focused, broadband THz beam through the sample. Transmission is consistent with a sheet conductivity $\sigma_s > 30\sigma_q = 30 \cdot e^2/4\hbar$, i.e., at least 30 times larger than the optical sheet conductivity associated with interband transitions in graphene [19]. Our measurements indicate that the optical response of graphene in the THz band is dominated by intraband transitions rather than interband transitions. The spectral response is flat, suggesting that our probe frequencies are well below the Drude roll-off frequency. By measuring THz transmission at discrete points across a graphene film we are able to map out sheet conductivity as a function of position. In contrast to conventional measurements of sheet conductivity, our THz imaging technique does not require patterning of graphene or fabrication of electrical contacts.

## 2. Experiment

Graphene was grown on Cu foil (Alfa Aesar, 25-μm thickness) at a temperature of 1000°C under a flow of methane (158 sccm) and hydrogen (6 sccm) inside a one-inch tube furnace [15]. To promote growth of single-layer graphene the pressure of the CVD system was reduced to 5.5 Torr [15]. To transfer the graphene from Cu to a Si wafer we used methods developed by previous authors [14,15]. High resistivity Si was used due to its high transparency to THz radiation. The top surface of the graphene-coated Cu foil was covered with PMMA (500nm thickness). The Cu was etched away by floating the sample in a $FeCl_3$-based solution (CE-200, Transene) leaving behind a PMMA-graphene film. The PMMA-graphene film was then washed in a series of six DI water baths and scooped out of solution using the Si substrate. The sample was allowed to air dry for 6 hours to promote adhesion of the graphene to the Si. Finally the PMMA was removed by a 6 hour soak in methylene chloride followed by 10 minute baths in acetone and IPA. After growth, the graphene on Si was verified to be predominately single-layer with low disorder by micro-Raman spectroscopy (Fig. 1) [20]. The position of the G-peak (1589 $cm^{-1}$) is consistent with a doping level $\sim 4 \times 10^{12}$ $cm^{-2}$ [21]. Extensive characterization of control samples was also performed using micro-Raman spectroscopy and scanning electron microscopy. We consistently find that the growth method produces > 90% coverage of single layer graphene.

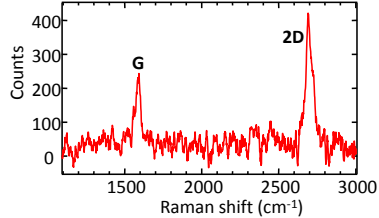

Fig. 1. Raman spectra from the graphene-on-Si sample. The background signal from bare Si has been subtracted. The integrated intensity of the 2D peak is more than double the G peak. There is no disorder-induced peak apparent at 1345 $cm^{-1}$. The same ratio of G peak to 2D peak was observed at multiple points across the graphene film.

We carried out two-dimensional raster scans of the graphene sample in a transmission geometry using broadband THz pulses. The broadband THz pulses were generated by optical rectification of femtosecond laser pulses in a 1-mm ZnTe crystal. The light source was a 1-kHz Ti:sapphire amplifier producing 800-nm femtosecond pulses (pulse energy, 1 mJ; pulse duration, 90 fs). The central frequency and the bandwidth of the THz pulses were 1 and 1.5 THz, respectively. THz pulses were focused onto the graphene/Si sample or the bare Si substrate. The beam size at the focus was 0.5 mm. The transmitted THz pulses were measured by either a L-He-cooled Si:Bolometer (sensitive to time-averaged THz power), or by electro-optic (EO) sampling using a 1-mm ZnTe crystal [22].

## 3. Power transmission: 2-D imaging and sheet conductivity calculation

Figure 2a shows transmitted THz power measured by the Si:Bolometer. The image covers a 26×41-$mm^2$ region. The pixel size is 0.4-mm and data was acquired with a 100-ms pixel integration time. The square-shaped graphene film (blue-green, average transmission: 0.39) is clearly resolved against the background of the Si substrate (bright-green, average transmission: 0.57). The THz response of the graphene film shows spatial inhomogeneity. The transmission varies from 0.36 (top right edge) to 0.41 (bottom left edge). We observed the transmission near the graphene edge in a 1.5×1.5-$mm^2$ region with 0.02-mm step size (Fig. 2b). The transmission drop across the boundary is as sharp as the spatial resolution of our probe, 0.5 mm (Fig. 2c).

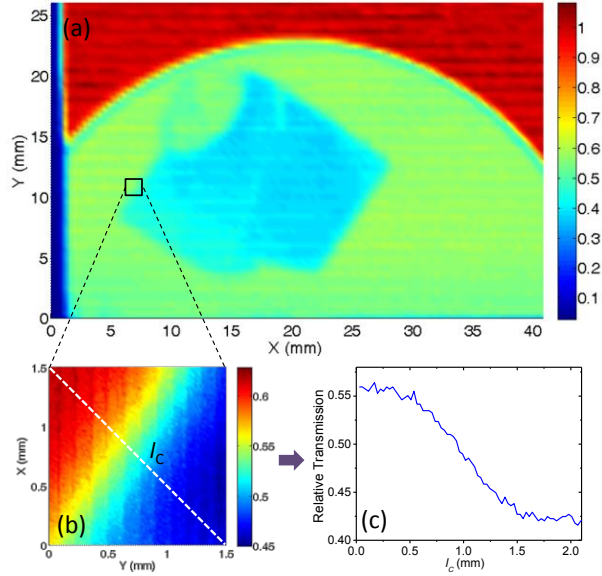

Fig. 2. (a) THz transmission image of the graphene-on-Si sample over a 26×41-mm² region (pixel size is 0.4-mm). The graphene film and the Si substrate are shown in light blue and bright green, respectively. The red and dark-blue regions correspond to air and the aluminum sample mount, respectively. Measurements were made at room temperature in ambient conditions. (b) A higher definition image (1.5×1.5-mm²) taken with 0.02-mm steps shows an edge of the graphene. (c) The cross-section of the edge is shown.

From the relative power transmitted through graphene/Si versus bare Si we calculated the sheet conductivity $\sigma_s$ of our single-layer graphene sample. The sample has a multi-layer structure consisting of graphene, Si, and air layers (Fig. 3) which we analyze using thin-film Fresnel coefficients and the Drude model. The graphene layer is treated as a zero-thickness conductive film, whereas the Si substrate is considered an optically thick dielectric medium. The high-resistivity Si substrate has refractive index $n_{Si} = 3.42$ and is nearly dispersionless in the THz regime. The transmission through the first interface (air→graphene→Si) is given by

$$t(\sigma_S) = \frac{2}{n_{Si} + 1 + Z_0 \sigma_S} \quad (1)$$

and the internal reflection from the graphene interface is given by

$$r(\sigma_S) = \frac{n_{Si} - 1 - Z_0 \sigma_S}{n_{Si} + 1 + Z_0 \sigma_S}, \quad (2)$$

where $Z_0$ (376.7 Ω) is the vacuum impedance. The ratio of the total transmission of the graphene/Si sample to that of the Si substrate is given by,

$$T_{rel}(\sigma_S) = \frac{T_{Gr-Si}}{T_{Si}} = \left|\frac{t(\sigma_s)}{t_{13}}\right|^2 \frac{1 - |r_{34} r_{31}|^2}{1 - |r_{34} r(\sigma_s)|^2} \quad (3)$$

where $t_{ij} = 2n_i/(n_i + n_j)$ and $r_{ij} = (n_i - n_j)/(n_i + n_j)$ are the Fresnel coefficients with the refractive indices of $n_1 = n_4 = n_{air} = 1$ and $n_3 = n_{Si} = 3.42$. There are no interference terms in Eq. 3 because multiple reflections are temporally separated (see Fig. 4a). The measured values of local $T_{rel}$ varied from 0.64 to 0.72 depending on position, from which we calculate the local sheet conductivity $\sigma_s(x,y) = 1.7 \times 10^{-3}$ to $2.4 \times 10^{-3}$ Ω$^{-1}$ ($\rho_s$ = 420 to 590 Ω/sq). These values are plotted in Fig. 3b. We speculate that spatial inhomogeneity in $\sigma_s$ is caused by variations in doping level [23]. Variations in doping likely occur during the graphene transfer process. A clearer understanding of the causes of spatial variations in conductivity, and ways to improve the uniformity of transferred graphene films, is a subject for future work.

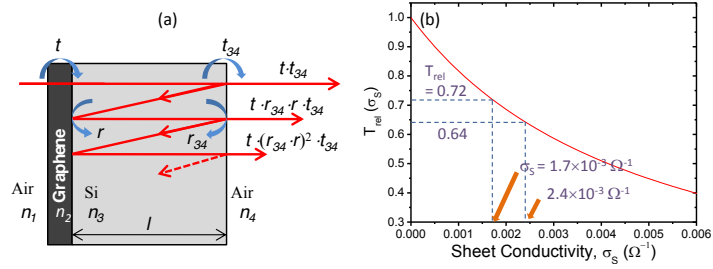

Fig. 3. (a) Multiple reflections at Air-Graphene-Si-Air interfaces: $n_1 = n_4 = n_{Air} = 1$ and $n_3 = n_{Si} = 3.42$. (b) Relative THz transmission of graphene/Si versus sheet conductivity

The measured sheet conductivity is more than 30 times greater than $\sigma_q = e^2/4\hbar = 6.1 \times 10^{-5}$ $\Omega^{-1}$ ($\sigma_q$ is the optical conductivity of graphene due to interband transitions [19]). We conclude that the measured sheet conductivity is dominated by intraband transitions and should closely reflect the dc electrical conductivity of the graphene sample. To compare our THz measurements of $\sigma_s$ to conventional techniques we patterned 200 µm van der Pauw squares in the graphene film. Four-probe dc electrical measurements of these patterned graphene films yielded $\rho_s$ ranging from 630 to 750 Ω/sq in reasonable agreement with the THz measurements. We propose three possible causes for the 30% increase in measured $\rho_s$, (i) grain-boundary scattering has a larger effect on the dc-electrical measurements than the THz measurements, (ii) small voids in the graphene film have a larger effect on the dc-electrical measurements than the THz measurements, (iii) the additional semiconductor processing steps required to pattern graphene and fabricate metal electrodes may reduce the doping level of the graphene, thereby increasing the measured $\rho_s$.

## 4. Terahertz time-domain spectroscopy and conductivity spectra

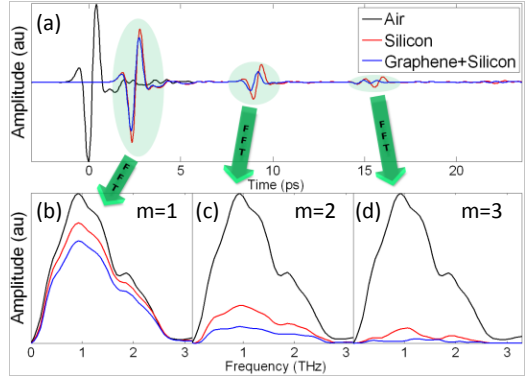

Fig. 4. (a) THz waveforms transmitted through air (black), silicon (red), and graphene on silicon (blue). (b-d) Amplitude spectra of the directly transmitted pulse (m=1) and the first two internally reflected pulses (m=2 and 3) through silicon (red) and graphene on silicon (blue). The spectrum of the pulse through air is added for comparison (Black line).

The reflection and transmission coefficients described in Eq. 1 and 2 can be determined as a function of frequency using THz time-domain spectroscopy (THz-TDS). Figure 4a shows a set of data including the THz waveforms through air (black), the Si substrate (red), and the graphene/Si sample (blue). The waveforms measured from both Si and graphene/Si consist of a series of single-cycle THz pulses. First, a directly transmitted pulse ($m = 1$), then subsequent "echos" corresponding to multiple reflections from the front and back sides of the Si substrate ($m = 2, 3, 4…$). The time delay between echoes is consistent with the thickness of the Si substrate (285±5 µm). The amplitude difference between graphene/Si pulses and Si pulses becomes more pronounced as the pulses undergo more reflections. Figure 4b-d shows Fourier transforms of the $m=1$, $m=2$ and $m=3$ waveforms respectively, obtained with a 6-ps time window corresponding to the time delay between echos. High-resolution Fourier transform of the entire waveform shows no sign of narrow spectral features other than the interference fringes of the periodic pulse train.

Combining Eqs. 1 and 2 with $t_{ij}$, and $r_{ij}$, the relative field transmission of the $m$-th pulse is predicted to be

$$t_{rel}^{(m)}(\sigma_S) = \frac{E_{G-Si}^{(m)}}{E_{Si}^{(m)}} = \frac{t(\sigma_S)}{t_{13}}\left(\frac{r(\sigma_S)}{r_{13}}\right)^{m-1}. \quad (4)$$

where $E_{G-Si}^{(m)}$ is the electric field of the $m$-th pulse after transmission through graphene/Si and $E_{Si}^{(m)}$ is the electric field of the $m$-th pulse after transmission through Si. Assuming $\sigma_s = 2.04\times10^{-3}$ $\Omega^{-1}$ (the spatially-averaged sheet conductivity of graphene obtained from the power transmission measurement), Eq. 4 predicts $t_{rel}^{(1)} = 0.852$, $t_{rel}^{(2)} = 0.495$, and $t_{rel}^{(3)} = 0.288$, in good agreement with the pulse-energy ratios ($t_{rel}^{(m=1,2,3)} = 0.855$, $0.454$, and $0.299$) seen in Fig. 4.

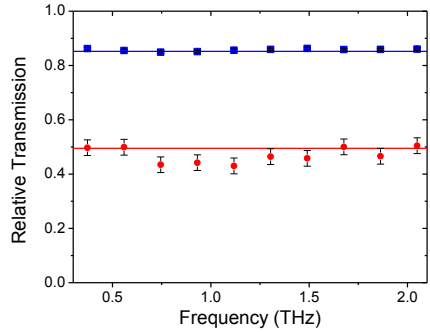

Fig. 5. Relative amplitude transmission spectra of $m=1$ (blue square) and $m=2$ (red circle) pulses (graphene-on-Si transmission spectra divided by the Si transmission spectra). Solid lines at $t_{rel} = 0.852$ and $0.495$ show the expected relative amplitude based on the spatial-average of our local sheet conductivity measurements $\sigma_s = 2.04\times10^{-3}$ $\Omega^{-1}$. The experimental spectra were obtained by averaging the transmission through five different spots on the graphene.

Figure 5 shows the relative transmission spectra through the graphene/Si sample for pulses $m=1$ and $m=2$ (i.e., transmission through graphene/Si relative to transmission through bare Si). The spectra are flat and in close agreement with the expected values $t_{rel}^{(1)} = 0.852$ and $t_{rel}^{(2)} = 0.495$. The flat spectral response seen in Fig. 5 indicates that the period of the applied electric field (0.5-3 ps) is much longer than the room-temperature carrier scattering time in our graphene sample [16,24].

### 5. Conclusion

We conclude that THz imaging and spectroscopy is of great use for rapidly characterizing the local free carrier dynamics in graphene. We have demonstrated that the strong THz absorption of graphene leads to high contrast imaging and the ability to accurately map sheet conductivity with sub-mm resolution over large areas.

**Acknowledgments:** The work at the Oregon State University is supported by Oregon Nanoscience and Microtechnologies Institute and by National Science Foundation (PHY-0449426). The work at Cornell was supported by the NSF through the Cornell Center for Materials Research (CCMR), the MARCO Focused Research Center on Materials, Structures, and Devices and the AFOSR. Sample Fabrication was performed at the Cornell node of the National Nanofabrication Infrastructure Network, which is supported by the National Science Foundation (Grant ECS-0335765).